\documentclass[aps,twocolumn,showpacs]{revtex4}
\usepackage[dvips]{graphicx}
\usepackage{amsmath}
\newcommand{\bea}{\begin{eqnarray}}
\newcommand{\eea}{\end{eqnarray}}
\newcommand{\be}{\begin{equation}}
\newcommand{\ee}{\end{equation}}

\newcommand{\ltwid}{\mathrel{\raise.3ex\hbox{$<$\kern-.75em\lower1ex\hbox{$\sim$}}}}
\newcommand{\gtwid}{\mathrel{\raise.3ex\hbox{$>$\kern-.75em\lower1ex\hbox{$\sim$}}}}

\begin{document}

\title{Interface-mediated pairing in field effect devices}

\author{V.~K\"orting$^{1}$, Qingshan Yuan$^{1,2,3}$, P.~J.~Hirschfeld$^{1,4}$,
T.~Kopp$^{1}$, and J.~Mannhart$^{1}$}

\affiliation{$^1$Center for Electronic Correlations and Magnetism,
EP6, Univ.  Augsburg, Augsburg Germany \\
$^2$Texas Center for Superconductivity and Advanced Materials,
Univ.\ of Houston, Houston, TX 77204 USA\\
$^3$Pohl Institute of Solid State Physics, Tongji University, Shanghai 200092,
P.R. China\\
$^4$Department of Physics,
University of Florida, PO Box 118440, Gainesville FL 32611 USA }
\date{\today}

\begin{abstract}
We consider the pairing induced in a strictly 2D electron gas
(2DEG) by a proximate insulating film with polarizable localized
excitations.  Within a model of interacting 2D electrons and
localized two-level systems, we calculate the critical temperature $T_c$
as a function of applied voltage and for different materials
properties. Assuming that a sufficient
carrier density can be induced in a field-gated device, we argue
that superconductivity may be observable in such systems.
$T_c$ is found to be a nonmonotonic function of both
electric field and
the excitation energy of the two-level systems.
\end{abstract}

\maketitle

Shortly after the publication of the BCS theory of
superconductivity, W.~A.~Little proposed a possible pairing
mechanism for electrons in long organic molecules involving
localized electronic excitations in the molecules' side
chains~\cite{Little}.
The suggested advantages of such an
arrangement included the large characteristic energy of the
couplings, which might lead to high temperature superconductivity
(HTSC), and the separation of the excitations themselves from the
screening effects of the electron gas. Such a
pair mechanism has never been realized, presumably due to large
fluctuation effects in quasi-1D systems. Later,
Ginzburg~\cite{Ginzburg} and Allender, Bray and Bardeen~\cite{ABB}
proposed a similar excitonic 2D mechanism in superconductor-semiconductor
sandwiches.
The enormous body of early work on this problem has
been reviewed in~\cite{Ginzburg2}.
One problem with
these schemes is clearly that the localized nature of the excitations
implies that at most one or two atomic layers of the intercalating
insulator can contribute to the pairing, meaning that the enhancement
of pairing in the relatively thick nearby superconductor is negligible.

Here we propose that a similar scheme might work for an insulating
layer in proximity with a superconducting layer of near-atomic
thickness.   Such systems can in principle be prepared in several
ways, but the most promising is perhaps the field effect
method pioneered in the early 90's with the intent of increasing
the carrier density and hence $T_c$ in  HTSC cuprate
devices~\cite{Hebard,mannhart91,mannhart96,mannhartreview, GennaLog}.
 This work has shown incontrovertibly that $T_c$ and
 other superconducting properties in metallic
samples can be influenced by an applied gate voltage.  While the
experiments have most often been interpreted in terms of
electronic structure variations near the interface, they are not
understood in detail.

It is interesting to ask if new physics can result from field
doping of insulators.  In the most recent reports on
field-doping of SrTiO$_3$, Pallecchi {\it et al.}~\cite{Pallecchi01}
and  Ueno {\it et al.}~\cite{Ueno03} were
able to achieve an areal carrier density of $\sim$~0.01--0.05 per unit cell.
It is difficult to achieve higher densities
due to electrical breakdown in the insulating layer at high fields.
But there seems to be no {\it fundamental} objection to
even higher charge densities, since complex oxide dielectrics and
ferroelectric oxides can achieve polarizations in the range of
$\sim 0.5$ ~\cite{mannhartreview}. In fact, very recently the first
observation of field-induced superconductivity was reported for a
device with a Nd$_{1.2}$Ba$_{1.8}$Cu$_3$O$_7$ epitaxial film grown on
SrTiO$_3$ substrates~\cite{Cassinese04}.

Therefore one may legitimately ask the question what critical
temperature can be achieved in a
system where the pairing comes entirely from the excitations
in the proximate insulating layer, and
how $T_c$ is likely to vary with field in this case.
Alternatively, one can assume the existence of
a 2D superconducting layer with preexisting pair
interaction and bare critical temperature $T_c^0$,
and ask by how much the presence of the insulating layer
enhances $T_c$. While in a strictly 2D system the $T_c$'s referred to cannot
correspond to true long-range order, the creation
of a field-tuned superconducting state with algebraic order at
finite temperatures would be of
considerable interest and applicability in small devices.

We begin by considering an insulating amorphous film $L1$
in proximity to a correlated insulator $L2$ (drain-source (DS) channel of a field
effect device), similar to Ref.~\onlinecite{Cassinese04}, with a small
density of localized charge carriers. Applying an electric field as shown in
Fig.~\ref{device} sweeps charge carriers to the interface with the
film $L1$ where they accumulate~\cite{comment7}.
The formation of the 2D band induced by an electric field
has been studied by Poilblanc {\it et al.}~\cite{Wehrli}.
We assume for the moment that
in the absence of the film $L1$ there are no pairing interactions
between the electrons in the material $L2$. Qualitatively, we expect the following
picture to apply: virtual excitations in the dielectric $L1$
induce Cooper pairing in the adjacent layer $L2$. The critical
temperature must increase initially with the field since carriers
are being injected into the system.
With increasing electric field
the larger level splitting of the two level system leads to a
suppression of the polarization fluctuations and the pair
potential decreases.
$T_c$ is therefore expected to reach a
maximum at a characteristic field strength; it is our
objective to estimate the scale of possible $T_c$'s through this process, as
well as the field strength required to attain it.

We now propose a crude but concrete framework within which one can
calculate these effects.   Roughly speaking, the dielectric can
have two effects on $T_c$. First, it can reduce the Coulomb
pseudopotential of electrons in the field-doped layer due to the
large dielectric constant of the amorphous insulator $L1$.
We are more concerned here with the
second effect, namely an additional contribution to the residual
pairing interaction at the interface due to virtual polarization
of the dielectric itself.

\section{model}
\begin{figure}[t]
\centering
\includegraphics[width=1\columnwidth,clip]{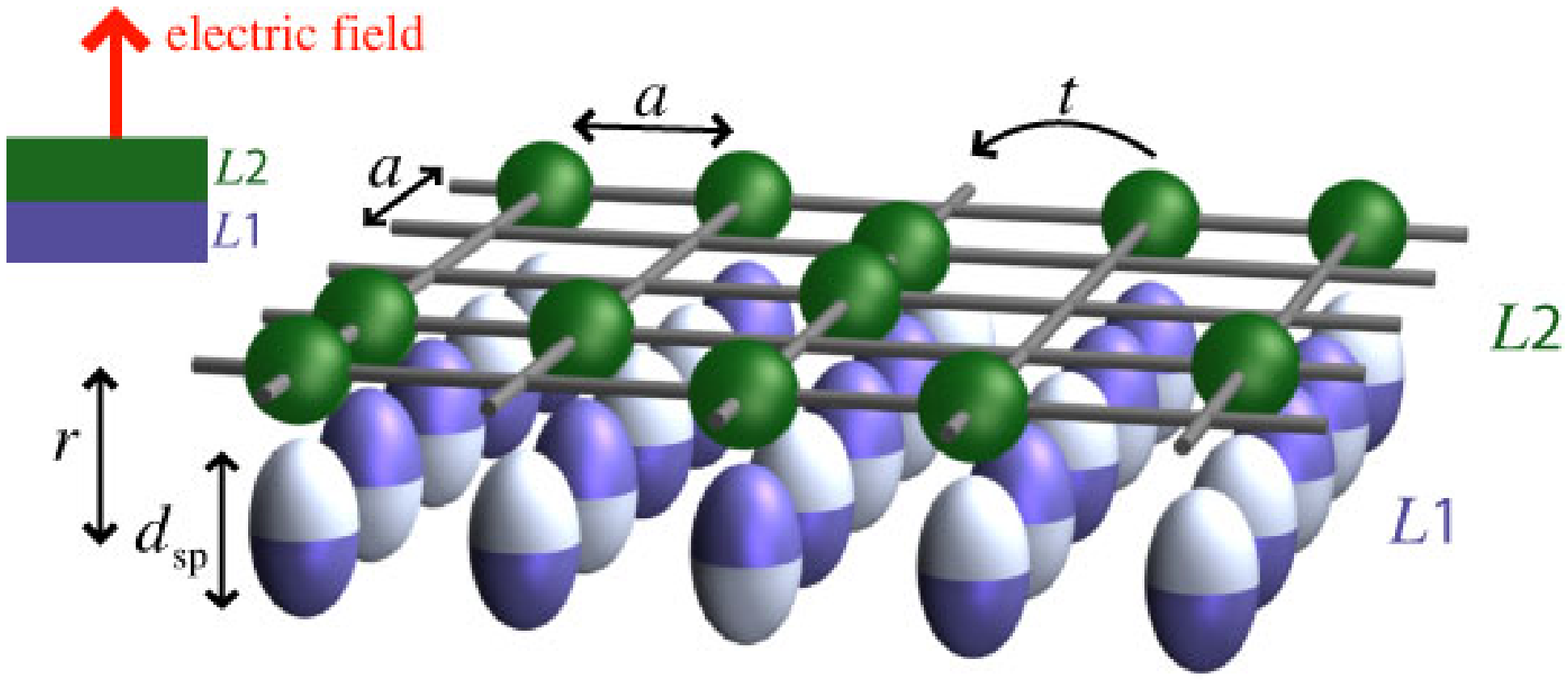}
\vspace{-4mm}
\caption{
Our model of a field effect transistor is represented by two layers ($L1$ and $L2$): \\
$L1$: dielectric layer with local dipoles of dipole moment $e d_{\rm sp}$,
presented by two-level systems; \\
$L2$: metallic DS-channel with a 2D electron gas.
}{\label{device} }
\end{figure}

 The Hamiltonian we consider describes a single layer $L2$ of electrons
 $c_i^\dagger$ hopping on a square lattice with lattice constant $a$, nearest neighbor
 hopping $t$, and a set of localized two-level systems in an adjacent
 layer $L1$  with ground state $s_i^\dagger$ and excited state
 $p_i^\dagger$ separated in energy by $\Delta_{sp}$, and their
 respective dipole moment is $ed_{sp}$ (Fig.~\ref{device}).
 Note these states simply designate ground and excited states of a
 localized two level system, and need not correspond to
 actual atomic s-- and p-- orbitals.
 To ensure that the two-level system is occupied by only one electron,
 the constraint $s_i^\dagger s_i + p_i^\dagger p_i=1$ must be enforced.
 The electric field
\begin{equation}\label{Efield}
{\cal E}\equiv {E_{sp}}/(e\, d_{sp})
\end{equation}
is assumed to populate the
 metallic layer and  simultaneously polarize the two level systems by driving transitions
 between the $s$ and $p$ states, which are coupled to
 the free electrons in the layer $L2$ by a contact interaction
$V_{sp}$.

The Hamiltonian contains, according to our assumptions, the following
elementary processes:
\begin{equation}
H_{tot}  =  H_{t} + H_{2l} + H_{ext} + H_{int} + H_{\mu}+ H_{e-e}
\end{equation}

with
\begin{eqnarray}
H_{t} & = & - \, t \sum_{ <ij>, \sigma}
               c_{i, \sigma}^{\dagger} c_{ j, \sigma}
      \label{Ht} \\
H_{2l} & = & \frac{1}{2} \Delta_{sp} \sum_{i}
              ( p_{i}^{\dagger} p_{i} - s_{i}^{\dagger} s_{i} )
      \label{Hlevel} \\
H_{ext} & = &  E_{sp} \sum_{i}
              ( p_{i}^{\dagger} s_{i} + s_{i}^{\dagger} p_{i} )
      \label{Hext} \\
H_{int} & = & V_{sp} \sum_{i, \sigma}
              c_{ i, \sigma}^{\dagger} c_{ i, \sigma}
              ( p_{i}^{\dagger} s_{i} + s_{i}^{\dagger} p_{i} )
      \label{Hint}
      \\
H_{\mu} & = & - \mu \sum_{  i, \sigma}
             c_{  i, \sigma}^{\dagger} c_{  i, \sigma}
    \label{Hmu}   \\
    H_{e-e} & = & U\sum_i c^\dagger_{i\uparrow} c_{i\uparrow}c^\dagger_{i\downarrow} c_{i\downarrow}\label{Hubbard}.
\end{eqnarray}
Here $H_t$ describes a band of noninteracting $2D$ electrons on a
square lattice, $H_{2l}$ the energies of the localized two level
system, $H_{ext}$ the coupling of the electric field to these
orbitals, $H_{int}$ the Coloumb interaction between electrons in
the metallic layer and the two level system, and $H_{\mu}$ the
chemical potential. The direct electron-electron interaction term
$H_{e-e}$ is taken to be local and repulsive. Very similar models
have been used recently to discuss dielectric properties of bulk
cuprates~\cite{vandenbrink}, the competition between  charge
density wave and superconductivity in two dimensional electronic
systems~\cite{su}, and superconductor-ferroelectric
multilayers~\cite{pavlenko}. The system is not exactly soluble,
but will be treated under the assumption that the polarization
through the charges in $L2$ is not large enough to drive the
system into the ferroelectric state. We first diagonalize
$H_{2l}+H_{ext}$, and then express the corresponding quasiparticle
operators in a pseudospin representation where $S_i^z$ measures
the occupation of the 2-level system and $S_i^\pm$ induces
transitions between the two eigenlevels. The occupation constraint
on the two-level system is preserved by the usual spin algebra.
The (exact) final form of the Hamiltonian is

\begin{eqnarray}
H_{2l} +H_{ ext} &=& - 2 \sqrt{ E_{sp}^{2}+(\frac{1}{2}
\Delta_{sp})^{2}}
                  \sum_{i} S_{i}^{z}
      \label{Hlevel_spin} \\
H_{int} &=& - \, V_{z} \sum_{ i, \sigma}
                       c_{ i, \sigma}^{\dagger} c_{ i, \sigma} S_{i}^{z}
      \label{Hint_Vz_spin} \\
        &&  + \, V_{x} \sum_{  i, \sigma}
                       c_{ i, \sigma}^{\dagger} c_{ i, \sigma}
                       (  S_{i}^{+} +  S_{i}^{-} )\quad ,
      \label{Hint_Vx_spin}
\end{eqnarray}
where
\begin{eqnarray}
  V_{z} &=& \, 2 V_{sp} \frac{E_{sp} }
          { \sqrt{ E_{sp} ^{2}+(\frac{1}{2} \Delta_{sp})^{2} } } \nonumber \\
  V_{x} &=&  \, V_{sp} \frac{ \frac{1}{2} \Delta_{sp}}
          { \sqrt{E_{sp} ^{2}+(\frac{1}{2} \Delta_{sp})^{2} } }\quad. \nonumber
\end{eqnarray}

The system of interacting spins-1/2 and fermions may now be
treated within linear spin wave theory, an approximation which is
justified {\it a posteriori} by the observation that the occupation of
the higher-energy 2-level state is very small,
due to the sufficiently large 2-level splitting~\cite{comment1}.
Introducing the usual
Holstein-Primakoff bosons~\cite{comment2}, we make the approximate replacements
$S_{j}^{+}\rightarrow  b_{j}$, $
  S_{j}^{-} \rightarrow  b_{j}^{\dagger} $, $
  S_{j}^{z} \rightarrow  \frac{1}{2} - b_{j}^{\dagger} b_{j}
$ to find

\begin{eqnarray}\label{Hsw}
H \!\! &=& - t\sum_{ <ij>, \sigma}
              c_{ i, \sigma}^{\dagger} c_{ j, \sigma}
       - \mu_r \sum_{  i, \sigma}
             c_{  i, \sigma}^{\dagger} c_{  i, \sigma}
 - E_{2l}\sum_{i} (\frac{1}{2} - b_{i}^{\dagger} b_{i})
 \nonumber\\&&
 + V_{z} \sum_{ i, \sigma}
                      c_{ i, \sigma}^{\dagger} c_{ i, \sigma}
                     b_{i}^{\dagger} b_{i}
        + V_{x} \sum_{ i, \sigma}
                      c_{ i, \sigma}^{\dagger} c_{ i, \sigma}
                     ( b_{i}  +  b_{i}^{\dagger} )
                     +H_{e-e}\nonumber\\~&&~
\end{eqnarray}
We have introduced the energy splitting of the two-level system
\begin{equation}
E_{2l} = 2 \sqrt{E_{sp}^{2}+(\frac{1}{2} \Delta_{sp})^{2}}
\end{equation}
and the chemical potential has been renormalized
$ \mu_r= \mu +V_z/2$.

We now declare that the Hamiltonian $H$ is as good a starting
point as (\ref{Ht})--(\ref{Hubbard}) provided $\langle
b_{j}^{\dagger} b_{j}\rangle \ll 1$. We therefore proceed to apply
a Feynman variational procedure to the exact $H$, Eq.~(\ref{Hsw}),
after a unitary transformation, and attempt to extract the pairing
mechanism. The dominant term for this mechanism is the term with
$V_x$ which, as in the phonon-induced superconductivity, produces
Cooper pairing in second order perturbation theory~\cite{Little}.
However, for the field strengths which we will consider,  the term
with $V_z$ is also not negligible and it will alter pairing in one
significant respect: since it modifies the 2-level splitting when
a charge carrier occupies the site coupled to the considered
2-level system, it will either increase the pairing interaction
(for reduced splitting, i.~e., negative $V_z$) or decrease pairing
(for enhanced splitting, i.~e., positive $V_z$)
--- an observation which may be derived straightforwardly from a mean field
decoupling. For the physical system that we consider, $E_{sp}$ and
$V_{sp}$ always have the same sign, and consequently  $V_z$ is
positive~\cite{comment3}. All of the above considerations can
also be derived in terms of states where charges in $L1$ are
localized in states of definite position with respect to the
interface, as depicted schematically in Fig. 1.  In this
representation, the $V_z$ interaction may be shown to correspond
to the repulsion of the electron in layer $L2$ and the
field-induced dipole in $L1$.

In order to incorporate this effect we will include below the
thermal average
$V_{z}\sum_{i, \sigma}\langle c_{i,\sigma}^{\dagger}c_{i,\sigma}\rangle
\, b_{i}^{\dagger} b_{i}=V_{z} n \sum_{i}b_{i}^{\dagger} b_{i}$ explicitly
in the energy of the two-level system before handling the bosonic
degrees of freedom. This is achieved by rewriting the fourth term of
Eq.~(\ref{Hsw})
\begin{eqnarray}\label{vz}
&V_{z}&\sum_{i, \sigma} c_{i,\sigma}^{\dagger}c_{i,\sigma}
b_{i}^{\dagger} b_{i}\nonumber\\
&& = -  N V_{z} n n_b + V_z n \sum_{i} b_{i}^{\dagger} b_{i}
+ V_z n_b \sum_{i, \sigma} c_{i,\sigma}^{\dagger}c_{i,\sigma}\nonumber\\
&& \quad +\; V_z \sum_{i, \sigma} \left[c_{i,\sigma}^{\dagger}c_{i,\sigma}-n/2\right]
                     \left[b_{i}^{\dagger} b_{i}-n_b\right]
\end{eqnarray}
where $n_b=\langle b_{i}^{\dagger} b_{i} \rangle$ is the number of
bosons (the relative number of inverted 2-level systems per site),
$n=(1/N)\sum_{i,\sigma}\langle c_{i,\sigma}^{\dagger}c_{i,\sigma}\rangle$ is
the charge carrier density in layer $L2$, and $N$ is the number of
sites in $L2$.
The second term on the right hand side accounts for a density-dependent
renormalization of the two-level splitting:
\begin{equation}\label{en}
E^\star_{2l} = E_{2l} + n V_z \, .
\end{equation}

In a first step we apply a Lang-Firsov (LF) transformation of the
Hamiltonian $H$ in order to identify the pairing interaction, and in
a second step we control the decoupling of the interaction terms
through Feynman's variational principle which also fixes the
parameters of the LF transformation.

The LF transformation of the Hamiltonian $\tilde H = U^{\dagger} H U$
is achieved by the following unitary operator $U$:
\begin{equation}
U = \exp \left[-S_1(\theta)\right]\, \exp \left[-S_2(\gamma)\right]
\end{equation}
with
\begin{eqnarray}\label{lf}
S_1(\theta) &=& -\frac{1}{2V_x}\;\theta\, \sum_i (b_i^\dagger -b_i)\\
S_2(\gamma) &=& \frac{V_x}{E^\star_{2l}}\;\gamma\,
         \sum_{i, \sigma} c_{i,\sigma}^{\dagger}c_{i,\sigma} (b_i^\dagger -b_i)
\end{eqnarray}
The parameters $\theta$ and $\gamma$ will be fixed through the
variation of the free energy. The standard form of the LF
transformation for the Holstein model takes $\theta=0$ and $\gamma=1$.
With a ``zero-phonon'' approximation
the (phononic) Holstein model accounts for exact results in
the antiadiabatic limit $E_{2l}/t\gg 1$.
The variation of the
parameters $\theta$ and $\gamma$ has been devised in order to
reproduce the adiabatic limit of the Holstein model as
well~\cite{zheng,fehske,roeder,yuan}; the static case is realized with
$\gamma\rightarrow0$ and with a finite $\theta$ which signifies a
displacement field. For the Holstein model, a second canonical
transformation of Bogoliubov type
$U _B= \exp \left[-\alpha\sum_i (b_i^\dagger b_i^\dagger+b_i
b_i)\right]$ is sometimes applied to allow for the anharmonicity
of the lattice fluctuations. For our model,
we have to suppress states with two bosons (corresponding to our
initial constraint for the 2-level system), and
subsequent variation of the free energy
shows that in fact $\alpha=0$.
Also the ``static displacement'' should be suppressed
since we do not intend to consider a finite static polarization.
We verified from the minimization that indeed $\theta\simeq0$ for the considered
range of small to intermediate $V_{sp}/4t$. For this reason, and to
simplify the notation, we fix $\theta$ and $\alpha$ to zero.

With the LF transformation $U = \exp \left[-S_2(\gamma)\right]$ we
obtain

\begin{equation}\label{Htilde}
\tilde H = E_0 +  H_{\rm LF} +  H_{V_x} +  H_{V_z} + H_{V_x V_z}
\end{equation}
where
$ E_0=-\frac{1}{2}NE_{2l} -NV_z\,n\,n_b $,  and
\begin{eqnarray}\label{Hlf}
H_{\rm LF}& = &   -  \, t
\sum_{ <ij>, \sigma}
            c_{ i, \sigma}^{\dagger}  c_{ j, \sigma}\;
                    e^{  \frac { V_{x} }{ E^\star_{2l} }\gamma ( b_{i}^{\dagger} - b_{i} ) }
                    e^{ -\frac { V_{x} }{ E^\star_{2l} }\gamma ( b_{j}^{\dagger} - b_{j} ) }
                          \nonumber\\
   && \;-\;  V_{\rm eff}\; \sum_{i}
                              c_{ i, \uparrow}^{\dagger} c_{ i, \uparrow}
                              c_{  i, \downarrow}^{\dagger}
                                 c_{  i, \downarrow}
                           \nonumber\\
   &&  -\mu_{\rm LF}\; \sum_{i, \sigma}
            c_{ i, \sigma}^{\dagger}c_{ i, \sigma}
            \,+\,E^\star_{2l}\;  \sum_{i}  b_{i}^{\dagger} b_{i}
\end{eqnarray}
with
\begin{eqnarray}\label{Hv}
H_{V_x} =&& (1-\gamma)\, V_x \sum_{i,\sigma}c_{ i, \sigma}^{\dagger}c_{ i, \sigma}
                       (b_i^\dagger + b_i) \label{H_V_x}\\
H_{V_z} =&& V_z \sum_{i,\sigma}\bigl (c_{ i, \sigma}^{\dagger}c_{ i, \sigma}-n/2\bigr)
                      \; \bigl( b_i^\dagger b_i -n_b\bigr)\\
H_{V_x V_z} =&&\!\!\!\!\!\! -\,\gamma\,  \frac{V_x V_z}{E^\star_{2l}}
               \sum_{i,\sigma,\sigma'}\bigl (c_{ i, \sigma}^{\dagger}c_{ i, \sigma}-n/2\bigr)
                   c_{ i, \sigma'}^{\dagger}c_{ i, \sigma'}
                       (b_i^\dagger + b_i) \nonumber \\
           && \label{H_V_x_V_z}
\end{eqnarray}
where the effective chemical potential is given by
\begin{eqnarray}
\mu_{\rm LF}=\mu &+& V_z/2 \,+\, \gamma(2-\gamma)\frac{V_x^2}{E^\star_{2l}}
   \nonumber \\
        &&  -\,(1-n)V_z\left(\gamma\frac{V_x}{E^\star_{2l}}\right)^2 -V_z n_b
\end{eqnarray}

\begin{figure}[t]
\centering
\includegraphics[width=0.9\columnwidth,clip]{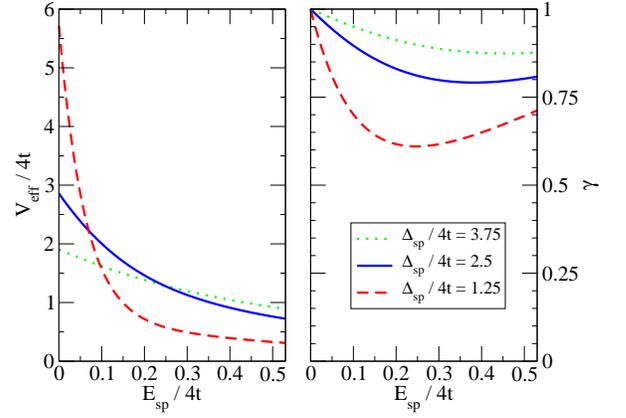}
 \caption{
Left panel: Effective pair interaction $V_{\rm eff}/4t$ vs.\
normalized electric field energy $E_{sp}/4t$ for various excitation
energies $\Delta_{sp}$ of the local two-level systems in layer $L1$.
The dielectric constant $\epsilon=100$ controls the increase of charge density in
the metallic layer $L2$ with electric field (cf.\ Eqs.~(\ref{nc})--(\ref{c})),
whereby we chose a bandwidth $4t= 400$~meV, a dipole length
$d_{sp}=2$~\AA\ (for charge transfer excitations), and a lattice constant
$a=4$~\AA. The coupling of the dipoles to the conduction
electrons is fixed at $V_{sp}/4t = 1.89$ which corresponds to a
spatial distance $r/a =1.5$ (cf.\ Eq.~(\ref{Vsp})).
Right panel: variational parameter $\gamma$ vs.\ normalized electric
field $E_{sp}/4t$, where $\gamma$ is found from the minimization of the free energy
(rhs of Eq.~(\ref{FeynmanVP})).}
{\label{gammaVeff}
}
\end{figure}

The induced electronic interaction $V_{\rm eff}$ is now
\begin{equation}\label{Veff}
V_{\rm eff} = 2 \,
\frac{V_x^2}{E^\star_{2l}}\gamma\left[(2-\gamma)
                       -\gamma\, (3-n)\frac{V_z}{E^\star_{2l}}\right] -
                       U\,,
\end{equation}
where we note that the Hubbard term $H_{e-e}$ has been unaffected
by the above transformations of the electron-two level system
coupling terms, due to the fact that it is simply a product of
local densities.  The local repulsion therefore merely diminishes
the effective attraction by $U$.

If $\gamma=1$, the first term within brackets in Eq.~(\ref{Veff})
coincides with the well-known result from the exact mapping of the Holstein
model to the attractive Hubbard model in the high-frequency limit
$(E_{2l} \rightarrow \infty)$. 
The  second term in the brackets $\propto V_z$
is always repulsive and suppresses $T_c$ for
increasing field. The electric field dependence
enters through $V_x$, $V_z$, $E^\star_{2l}$ and the implicit
dependence of $\gamma$ on $E_{sp}$.

We have now achieved the desired result of expressing the Hamiltonian
in terms of an effective BCS-like  interaction between electrons,
at the cost of introducing bosonic phase factors into the hopping
matrix elements.

We begin by  setting $U=0$ and investigate the
magnitude of the  attractive interaction obtainable by polarizing
the two level systems. In Fig.~\ref{gammaVeff} we illustrate how
$V_{\rm eff}$ depends on applied electric field. The main point is
physically obvious: if the electric field is sufficiently large,
it polarizes the electric dipoles in $L1$ and suppresses the
Little-type mechanism. We also note that if the bare excitation
energy $\Delta_{sp}$ of the dipoles in the insulating layer is
small, the intrinsic low-field pairing strength can be quite
large, up to several electron volts. However $\Delta_{sp}$ should
not be set to considerably smaller values than in
Fig.~\ref{gammaVeff} in order to guarantee the condition that the
occupation of the excited state is negligible.

In the second step we introduce an exactly solvable test Hamiltonian $H_{\rm test}$
and determine the fields in $H_{\rm test}$ through a Bogoliubov inequality
for the free energy $F$ of the model system~\cite{feynman}:
    \begin{equation}\label{FeynmanVP}
   F \leq F_{\rm test} + \langle \tilde H - H_{\rm test}\rangle_{\rm test}
    \end{equation}
where $\langle\;\rangle_{\rm test} $ signifies the thermodynamic
average with the test Hamiltonian.
We already anticipate the result of this variational scheme~\cite{comment4} and write
\begin{eqnarray}
      H_{\rm test} &=& E_{0,{\rm test}} - \, t_{\rm eff}
                       \sum_{<ij>, \sigma}  c_{i, \sigma}^{\dagger} c_{j, \sigma}
                   \; -\; \mu_{\rm LF} \sum_{i} c_{i, \sigma}^{\dagger} c_{i,\sigma}
                                       \nonumber \\
         &&+\;  \biggl(\Delta\,\sum_{i}
              c_{i, \uparrow}^{\dagger} c_{i, \downarrow}^{\dagger}+
                       h.c.\biggr)
                  \, - \, \Delta  \sum_{i}  \langle
                     c_{i, \uparrow}^{\dagger} c_{i,
                       \downarrow}^{\dagger} \rangle
               \nonumber \\
          &&
           \;+\;  E^\star_{2l} \sum_{i} b_{i}^{\dagger} b_{i} \label{H_test}
\end{eqnarray}
with
\begin{eqnarray}
        E_{0,{\rm test}} &=& E_0 -
          \frac{N}{4}\,n^2\,V_{\rm eff}\\
        t_{\rm eff} &=& t \, \exp
                 \left[ -
                \left( \frac{V_{x}}{E^\star_{2l}} \, \gamma \right)^{2}
             \coth \left( \frac{\beta\,E^\star_{2l}}{2}
        \right)  \right]
    \end{eqnarray}
We have assumed $s$-wave pairing for simplicity.  The associated
gap equation then reads
\begin{equation}
\Delta = V_{\rm eff} \sum_{\bf k}  {\Delta\over 2 E_{\bf k}}\tanh
{E_{\bf k}\over 2 T}
\end{equation}
and the Bogoliubov quasiparticle dispersion
\begin{eqnarray}
E_{\bf k} &=& \sqrt{\xi_{\bf k}^{2} + \Delta^{2} } \\
\xi_{\bf k}  &=& \varepsilon_{\bf k} - \mu_{\rm LF} \nonumber\\
    \varepsilon_{\bf k} &=& - 2\,t_{\rm eff} \, ( \cos k_{x} + \cos k_{y})
    \nonumber
\end{eqnarray}

\begin{figure}[t] \centering
\includegraphics[width=0.9\columnwidth,clip]{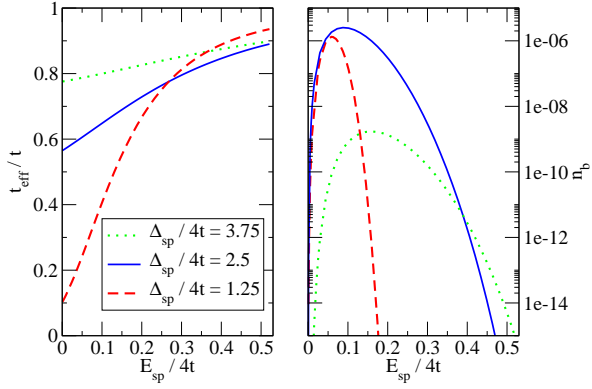}
\caption{
Left panel:
Effective hopping $t_{\rm eff}/t$  vs.\
normalized electric field energy $E_{sp}/4t$ at the transition temperature
$T_c$; all paramaters are identical to those of Fig.~\ref{gammaVeff}.
Right panel:
bosonic occupation number $n_b$ (the relative number of inverted
two-level systems per site) vs. normalized electric field energy $E_{\rm sp}/4t$
at the transistion temperature $T_c$.}
{\label{teff_nb} }
\end{figure}

To understand qualitatively the physics of the correlations
induced by interaction of the metallic layer with the 2-level
systems in $L1$, we display in the left panel of Fig.~\ref{teff_nb} the magnitude of
the renormalized hopping $t_{\rm eff}$. The band narrowing can be
rather significant for small excitation energies $\Delta_{sp}$
but for the parameters of greatest interest will turn out to be
only a factor of 1--5.

The fields still depend on the parameter $\gamma$;
correspondingly the rhs of relation Eq.~(\ref{FeynmanVP})
has to be minimized with respect to $\gamma$. For this purpose
we need the explicit form of $F_{\rm test}$:
\begin{equation}
F_{\rm test}=E_{0,{\rm test}}+\frac{1}{\beta}\,N\,\ln\left(1-e^{-\beta\,E^\star_{2l}}\right)
          + F_{\rm BCS}
\end{equation}
with
\begin{eqnarray}
F_{\rm BCS}\!\! &=& -\,\frac{2}{\beta}\sum_{\bf k}
                  \ln \left(1+e^{-\beta\,E_{\bf k}}\right)
              \nonumber\\
                && - \sum_{\bf k}
            \Bigl(
         E_{\bf k} -\xi_{\bf k}+\Delta\,\langle c_{-k,\downarrow}c_{k,\uparrow}\rangle
         \Bigr)
\end{eqnarray}

Since we have chosen $E_{0,{\rm test}}$ so as to guarantee the relation
$\langle \tilde H - H_{\rm test}\rangle_{\rm test}=0$, the
variation reduces to finding the optimal value of $\gamma$ from
the  minimization of $F_{\rm test}$:

\begin{equation}\label{gamma}
      \gamma = \frac{ n (n + 2) }
                   {
                    n (n + 2)
                    \,+\, n (n + 1) (2 - n) ({V_z}/{E^\star_{2l}})
                    \;+\;\delta(\gamma)
                   }
 \end{equation}
where

\begin{equation}
\delta(\gamma) \;= \; 2 \frac{\bar \varepsilon}{E^\star_{2l}}
                         \coth \big( \frac{ \beta E^\star_{2l}}{ 2 }\big)
\end{equation}

and

\begin{equation}
     \bar{ \varepsilon } = \sum_k \frac{\varepsilon_k}{ 1 + e^{\beta \xi_k} }
\end{equation}
For the range of parameters considered in our evaluation the function
$\delta(\gamma)$ can be neglected with respect to the other terms in
the denominator of Eq.~(\ref{gamma}). This allows us to calculate $\gamma$
algebraically from  Eq.~(\ref{gamma}). The validity of this
approximation has been proved by iterating the implicit Eq.~(\ref{gamma}).
Fig.~\ref{gammaVeff} (right panel) displays the decrease of
$\gamma$ with increasing $E_{sp}$ for various values of $\Delta_{sp}$ at
fixed $V_{sp}/4t=1.89$.

Finally the assumption below Eq.~(\ref{Hint_Vx_spin}) that the bosonic
occupation number $n_b = \langle b^\dagger_i b_i \rangle$ is very small
is in fact justified a posteriori for temperatures up to the
transition temperature $T_c$.
In the considered parameter range of the two-level splitting
$\Delta_{\rm sp}$ and of the applied gate field $E_{\rm sp}$,
$n_b$ is always smaller than $10^{-5}$.
Typical field dependences of $n_b$ are displayed in the
right panel of Fig.~\ref{teff_nb}.

\begin{figure}[h] \centering
\includegraphics[width=0.9\columnwidth,clip]{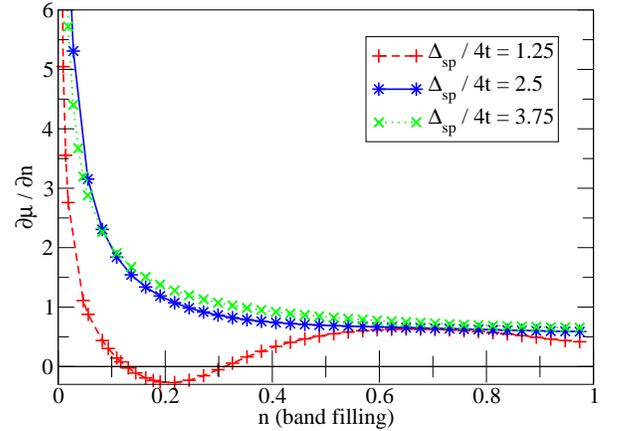}
\caption{
The derivative of the chemical potential with respect to particle
density, ${\partial\mu}/{\partial n}$, versus particle density $n$
at the respective $T_c$, calculated in Sec.~II. All paramaters are
identical to those of Fig.~\ref{gammaVeff}.
For a positive derivative, the normal and superconducting states are
globally stable in the vicinity of the transition.
}
{\label{mu_density} }
\end{figure}

The effective model does not implement the long-range part of the Coulomb
interaction. A model with purely local interactions, however, has a tendency
towards phase separation. In order to investigate if the transition from the
normal to the superconducting state is a transition between global minima of
the free energy, we evaluated the derivative of the chemical potential with
respect to carrier density. In this work, we focus on the transition into
the superconducting state and not on the evolution of the superconducting
state at lower temperatures. Then it is sufficient to discuss the
stability in the vicinity of $T_c$. Fig.~\ref{mu_density} illustrates
our conclusion that the derivative of the chemical potential with
respect to particle density is always positive for
$\Delta_{\rm sp}/4t$ larger than approximately $1.4$. Correspondingly,
the transition into the superconducting state is not preempted by a
competing transition into a phase separated state. At somewhat lower values of
$\Delta_{\rm sp}/4t$, the phase separation will probably be suppressed by
the non-local part of the Coulomb interaction.
As discussed in the following section, the relevant values of
$\Delta_{\rm sp} / 4t$ for a sufficiently strong
interface-mediated pairing appear in the regime where
${\partial\mu}/{\partial n}$ is positive.

\section{results}
In order to present the numerical results in the physically
relevant range of control parameters, we have to relate
microscopic quantities to laboratory parameters, such as the dielectric
constant $\epsilon$ of the insulating layer, the dipole length $d_{sp}$
of the two-level systems, and the distance $r$ from the two-level systems
to the sites of the conducting layer. Furthermore we have to establish
a relation between the electric field energy $E_{sp}$ (see Eq.~(\ref{Efield}))
and the number of charge carriers in $L2$. In our calculation
we assume that the charge carrier density is a
linear function of the electric field
with a field independent capacitance $C$ of
the dielectric $L1$: $Q=CV$, where $V$ is the voltage drop across the
dielectric and $Q$ is the total, accumulated charge at the interface
in $L2$. Then, the charge per square unit cell is
\begin{equation}
n=C\,\frac{{\cal E} d}{e}\,\frac{a^2}{A}
\end{equation}
where $a$ is the lattice constant, $A$ is the area of $L2$ and
$d$ its thickness (cf.\ Fig.~\ref{device}). With
$C=\epsilon_o\epsilon A/d$ and  Eq.~(\ref{Efield}) we
establish
\begin{equation}\label{nc}
n\,=\, c\;\frac{E_{sp}}{4t}
\end{equation}
with
\begin{equation}\label{c}
c\,=\, \epsilon_o\epsilon\,\, \frac{a^2}{e^2}\,\frac{4t}{d_{sp}}
\end{equation}

The interaction energy between the dipoles next to the interface
and the electric field of a charge carrier on the nearest site is
\begin{equation}\label{Vsp}
V_{sp}\;=\; \frac{1}{4 \pi \epsilon_0} \frac{e^2 d_{sp}}{r^2}\, .
\end{equation}

We note that dynamical screening of the interaction $V_{sp}$ should be included
in order to get more precise estimates. Although this will reduce somewhat
the high values of $T_c$ in our evaluation, it will not
alter the qualitative behavior of the $T_c$-dependence on the
gate field (see below).

Finally, the excitation energy of the dipoles has to be identified.
A generic dielectric is  not composed of a single species of
well-defined 2-level systems. Dipole excitations at various energies
are always present. We now address excitations in three different
energy ranges: for 10~eV (low atomic excitations with dipole length
$d_{sp}\simeq 1$~\AA), in the 1~eV range (charge transfer excitations
with $d_{sp}\simeq 2$~\AA), and in the meV range (ionic displacement
in atomic clusters with  $d_{sp}\simeq 0.1$~\AA ).

\subsection{Ionic displacement}
\begin{figure}[t]
\centering
\includegraphics[width=1.0\columnwidth,clip]{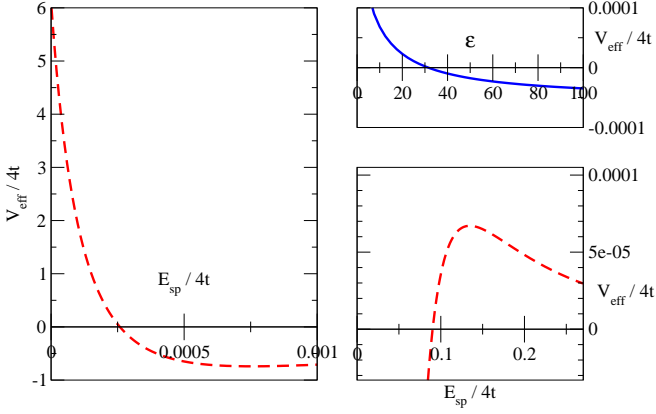}
\vspace{-2mm} \caption[]{\label{Veff_epsilon} Ionic displacement
in atomic clusters: induced  pairing interaction $V_{\rm eff}/4t$
versus electric field energy $E_{sp}/4t$ for $\epsilon=10$,
$d_{sp}= 0.1$~\AA\ and $r/a=1$ (which corresponds to
$V_{sp}/4t=0.2125$).  Left and lower right panels show
different scales. Upper right panel: Pairing interaction versus
dielectric constant $\epsilon$ for charge density $n=0.5$.}
\end{figure}

In dielectrics a displacement of ions, well localized at atomic
positions, usually accounts for the high dielectric constant. Such
displacements in atomic clusters typically correspond to a dipole
length of the order of $d_{sp}\simeq 0.1$~\AA\ and a small
excitation energy. In this case, the excitation energy is related
to $\epsilon$ and $d_{sp}$ through $\Delta_{sp}= 8\pi \,e^2
d_{sp}^2/{(a^3\,(\epsilon -1))}$. This relation follows directly
from the polarization density $\langle P \rangle$ of the
dielectric (with a cubic unit cell of volume $a^3$) in linear
response: $\langle P \rangle =  e d_{sp}\,
    \langle s_i^\dagger p_i + p_i^\dagger s_i \rangle/{a^3}
 = ({2 \,e^2 d_{sp}^2}/{a^3\,\Delta_{sp}})\, {\cal E}$
where $\langle P \rangle ={(\epsilon -1)/4\pi\,\cal E}$ holds.
Ionic displacements may have energies of the
order of 10~meV (for $\epsilon$ of the order of 20). However,
the induced pair interaction is far too small (see
Fig.~\ref{Veff_epsilon}) to find sizable transition
temperatures.
It is the repulsive term in
$V_{\rm eff}/4t$ of Eq.~(\ref{Veff}) which very effectively
impedes a transition to the superconducting state for such small
values of $\Delta_{sp}$.

\subsection{Charge transfer excitations}
In order to realize a sizable $T_c$ of the order of 100~K, excitations
at intermediate energy  and sufficiently large dipole length
have to be available for polarization. Charge transfer excitations
in the dielectrics are found at these energies --- e.g.\ for the transition metal (TM) oxides at
energies of the order of 1~eV, the charge transfer gap. Although these
excitations of, for example, a TM-oxygen plaquette or octahedron
become delocalized through hybridization,
we assume that most states near the interface have been localized by
disorder and local strain~\cite{comment5}. Furthermore
localized bound states below the charge transfer gap are conceivable.
It is not the focus of our present investigation to identify these
local excitations for a specific material. We rather point out
that such processes are possible and then evaluate $T_c$ as a function
of their respective energy.

We intend to focus on this latter type of localized charge
transfer excitations. For the evaluation we now fix the following
parameters: the bandwidth, the dipole length, and the lattice
constant:
\begin{eqnarray} \label{params}
4 t  &=&  400\, {\rm meV}  \nonumber \\
d_{sp}  &=& 2\, {\rm \AA}  \nonumber \\
a  &=&  4\, {\rm \AA}
\end{eqnarray}
Here, we chose a bandwidth and lattice constant typical of the
high-$T_c$ cuprates and many other oxides, and we take a dipole
length which corresponds to the interatomic oxygen-TM distance.
For excitation energies $\Delta_{sp}$ of more than 0.2~eV, the
primary polarization processes are electronic in nature.

\begin{figure}[t]
\centering
\includegraphics[width=1.0\columnwidth,clip]{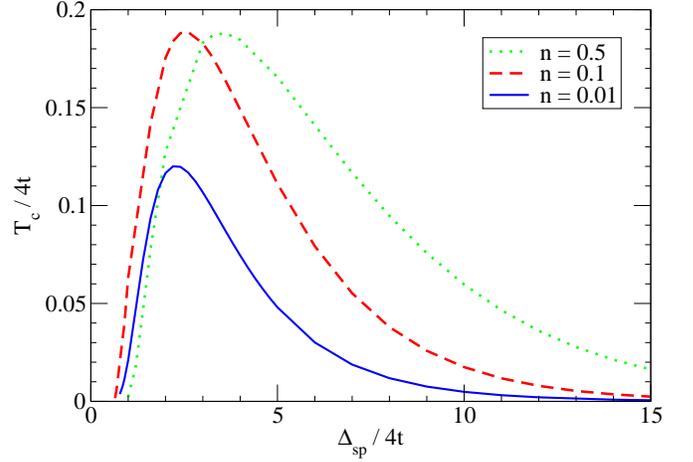}
\vspace{-2mm} \caption[]{\label{Tc_Deltasp}Charge transfer
excitations: transition temperature
$T_c/4t$ versus excitation energy $\Delta_{sp}/4t$.
The three curves present $T_c$ for small and intermediate band
filling. The dielectric constant
is $\epsilon=100$, which implies $c=1.87$ for the considered set
of parameters (see Eq.~(\ref{params})). The coupling of the dipoles to
the conduction electrons is fixed to $V_{sp}/4t = 1.89$ which
corresponds to a spatial
distance $r/a =1.5$ (cf.\ Eq.~(\ref{Vsp})). }
\end{figure}

In Fig.~\ref{Tc_Deltasp} the transition temperature to the
superconducting state is displayed as a function of the excitation
energy. In the range of small, increasing values of $\Delta_{sp}$
we observe a strong enhancement of $T_c$ whereas the transition
temperature decreases with excitation energies $\Delta_{sp}/4t$ above $\approx 2.5$.
The latter observation is expected since the pairing interaction
$V_{\rm eff}$ is inversely proportional to the excitation energy for
large $\Delta_{sp}$. For small $\Delta_{sp}$, the repulsive term
(second term in $V_{\rm eff}$, cf.\ Eq.~(\ref{Veff})) dominates and
suppresses the transition to the superconducting state for a finite
value of $\Delta_{sp}$. Note that the finite longitudinal
pseudospin-charge $V_z$ is decisive for the decay of $T_c$
at small excitation energies, outside the regime where the Holstein
model is appropriate.

\begin{figure}[t]
\centering
\includegraphics[width=1.0\columnwidth,clip]{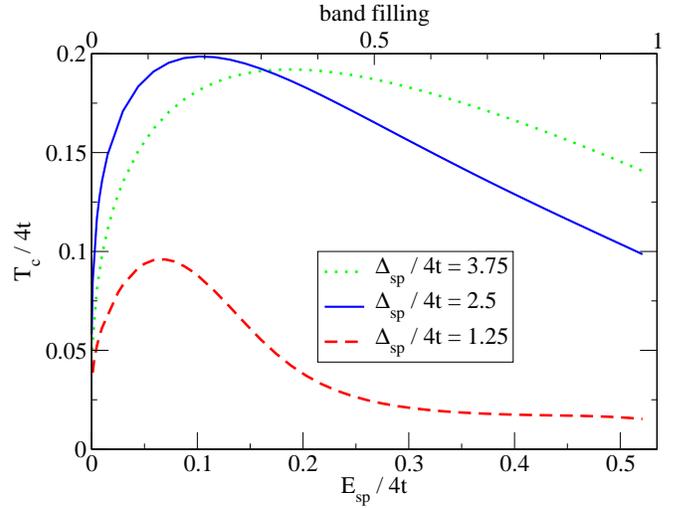}
\vspace{0mm} \caption[]{\label{Tc_Esp}  Charge transfer
excitations: transition temperature $T_c/4t$ versus electric field
energy $E_{sp}/4t$ and band filling $n$, lower axis and upper
axis, respectively. The three curves present different dielectrics
which are characterized by different excitation energies only,
other parameters are fixed for comparison. The dielectric constant
$\epsilon$ is 100 and $V_{sp}/4t = 1.89$ ($r/a = 1.5$). }
\end{figure}

For small increasing electric field, the transition temperature is
raised due to the accumulation of charge in the metallic layer
(cf.~Fig.~\ref{Tc_Esp}).  The electric field strength is
directly related to the band filling or induced areal charge
density. Strong electric fields with sizable band filling lower
the transition temperature as the repulsive term in $V_{\rm eff}$ is
enhanced and, moreover, the effective level splitting is
enlarged. In fact, as seen from the $\Delta_{sp}/4t = 1.25$
curve in Fig.~\ref{Tc_Esp} there are two scales for the
suppression of $T_c$ at higher fields. The lower scale is set by
the repulsive term in $V_{\rm eff}$ and is also responsible for
the observed decay of $T_c$ in the two further curves (with
$\Delta_{sp}/4t=2.5$ and 3.75). In this regime, $T_c$ is
being suppressed primarily by the increasing repulsion $V_z$
between the polarized dipoles and the 2D electrons, which scales
with the applied field. The larger scale, which is responsible for
the slow decay at even higher fields, is set by $\Delta_{sp}/4t$,
and corresponds to the eventual saturation of the dipole moment
of the  2-level systems.

The nonmonotonic dependence on filling or electric field is
reflected in the varying height of the three different curves in
Fig.~\ref{Tc_Deltasp}. However more striking is the small
variation of the position of the maxima in Fig.~\ref{Tc_Deltasp}
with filling (from 0.01 to 0.5). A value of $\Delta_{sp}\simeq
1$~eV seems to be optimal for a bandwidth of $4 t =  400$~meV.
This optimal excitation energy in units of the band width is
approximately
\begin{equation} \label{Delta_opt}
\Delta_{sp}^{\rm opt}/4t\,\simeq\, 2.5
\end{equation}

\begin{figure}[t]
\centering
\includegraphics[width=1.0\columnwidth,clip]{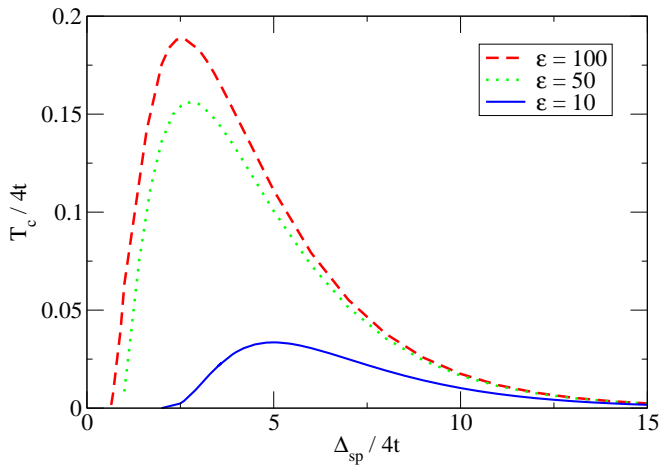}
\vspace{0mm} \caption[]{\label{Tc_epsilon}Charge transfer
excitations:  transition temperature $T_c/4t$ versus excitation
energy $\Delta_{sp}/4t$. The three curves present different
dielectrics ($\epsilon=10,50,100$). The band filling is fixed at
$n=0.1$, and the interaction  $V_{sp}/4t = 1.89$ corresponds to
$r/a=1.5$. }
\end{figure}

$\Delta_{sp}^{\rm opt}$ is weakly dependent on the parameters
$V_{sp}$, $d_{sp}$ and $\epsilon$ which mostly influence the
maximum value of $T_c$. For fixed charge density, dielectrics with
larger $\epsilon$ display a higher transition temperature (for
given $\Delta_{sp}$), see Fig.~\ref{Tc_epsilon}, as the electric
field necessary to create the charge density is smaller,
accounting for a larger $V_{\rm eff}$.

\begin{figure}[t]
\centering
\includegraphics[width=1.0\columnwidth,clip]{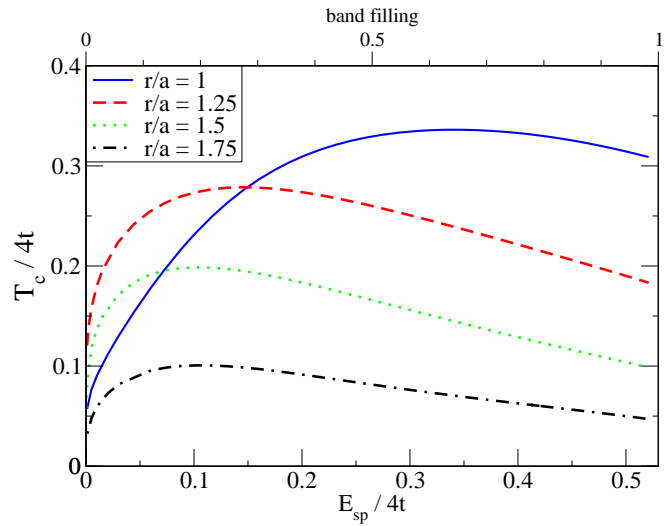}
\vspace{0mm} \caption[]{\label{Tc_cluster} Charge transfer
excitations ($d_{sp}=2$~\AA, $\Delta_{sp}/4t=2.5$, $\epsilon = 100$): transition
temperature $T_c/4t$ versus electric field energy $E_{sp}/4t$,
parameterized by $r/a$.  The corresponding $V_{sp}$ is in the
intermediate  coupling range: $V_{sp}/4t=4.25$ (corresponds to
$r/a=1$), $V_{sp}/4t=2.72$ ($r/a=1.25$), $V_{sp}/4t=1.89$
($r/a=1.5$), $V_{sp}/4t=1.39$ ($r/a=1.75$). }
\end{figure}

An increasing dipole charge-carrier interaction $V_{sp}$ is not only responsible
for an enhancement of $T_c$ but, for sufficiently large $V_{sp}$,
also for a ``retarded'' initial increase of $T_c$ with electric field (see Fig.~\ref{Tc_cluster}).
Again, this observation may be traced back to the field and
interaction dependence of the repulsive term in $V_{\rm eff}$.

For direct comparison with experimental devices, we note that
the gate voltage is related to the electric field energy plotted
in the figures through $E_{sp}/4t =e d_{sp} V/d$ where $d$ is the
thickness of the dielectric gate.
As a concrete example for a dielectric gate layer, we take SrTiO${}_3$.
Breakdown fields of 4$\times 10^7$~V/m  are reported for this
material with a low-$T$ dielectric function of order 100~\cite{christen}.
The maximum of the $T_c$ versus electric field energy curves occurs
for $n=0.2$ at about 2$\times 10^8$~V/m.
Thus the required fields for the {\it maximum} $T_c$ are only about
5 times higher than those already realized in this system.
Given that breakdown occurs due to ``pinhole"-type defects in the films,
it seems to us that the manufacture of samples with the required
breakdown fields is challenging but far from impossible.

\begin{figure}[ht]
\centering
\includegraphics[width=1.0\columnwidth,clip]{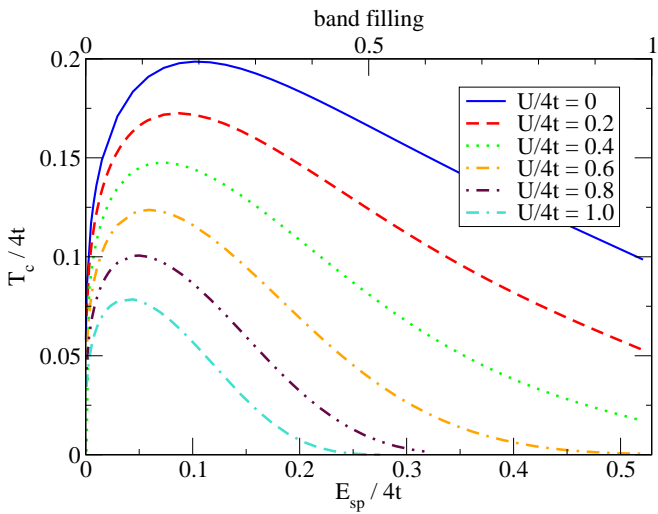}
\vspace{0mm} \caption[]{\label{finite_U}
Charge transfer
excitations: transition temperature $T_c/4t$ versus electric field
energy $E_{sp}/4t$ and band filling $n$, lower axis and upper
axis, respectively. A local electronic interaction $U$  (within layer
$L2$) is included in a weak coupling evaluation. The dielectric constant
$\epsilon$ is 100 and $V_{sp}/4t = 1.89$ ($r/a = 1.5$). }
\end{figure}

We conclude this subsection with a brief discussion of the
consequences that a nonzero local interaction $U$ within the
charge-carrier layer $L2$ has on the reduction of $T_c$ in a
weak-coupling evaluation. As shown in Eq. (\ref{Veff}), such a
repulsive Hubbard-type interaction does not modify the field
dependence of the effective interaction $V_{\rm eff}$,  but just
adds a constant $-U$. For weak interaction
$U/4t\ltwid 1$, the effective attraction $V_{\rm eff}$ is reduced
but not fully suppressed for small to intermediate fields (cf.\
Fig.~\ref{gammaVeff}) and corresponding filling. This observation
is reflected in the field and filling dependence of $T_c$ (see
Fig.~\ref{finite_U}): $T_c$ is reduced for small to intermediate
fields and suppressed for strong field-induced doping.
As a consequence of the field-dependent reduction of $T_c$, the maxima
are shifted towards lower values of filling, from about $n=0.2$ at
$U=0$ to $n=0.1$ at $U/4t=1$. The required fields for the {\it
maximum} $T_c$ are  reduced similarly.

\subsection{Atomic excitations}

\begin{figure}[ht]
\centering
\includegraphics[width=1.0\columnwidth,clip]{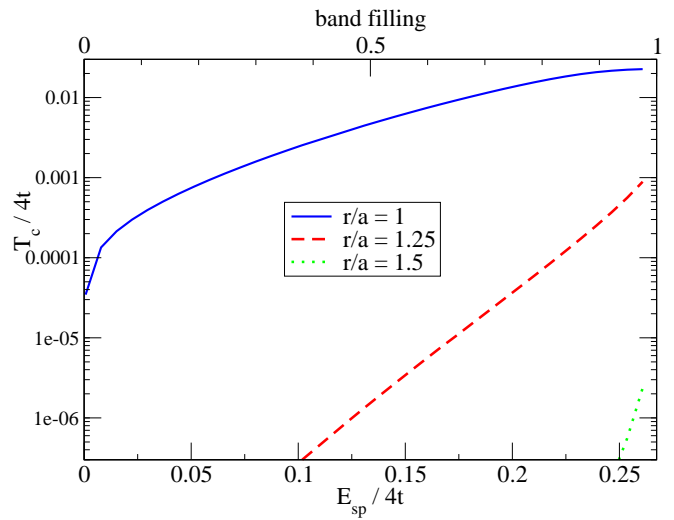}
\vspace{0mm} \caption[]{\label{Tc_atomic}
Atomic excitations ($d_{sp}=1$~{\rm \AA}, $\Delta_{sp}=10$~eV, $\epsilon = 100$):
transition temperature $T_c/4t$ versus electric field energy
$E_{sp}/4t$, parameterized by $r/a$:
$V_{sp}/4t=2.125$ (corresponds to $r/a=1$), $V_{sp}/4t=1.36$
($r/a=1.25$), $V_{sp}/4t=0.94$ ($r/a=1.5$).}
\end{figure}

At high energies, we encounter atomic excitations, such as
the $2p$ to $3s$ transition of O${}^{2-}$ states in the metal
oxides. These states are treated in exactly the same way in
the current theory, but are characterized by small dipole
lengths $d_{sp}\sim 1$\AA\ and large 2-level splittings
$\Delta_{sp}\sim$10 eV. Polarizing such dipoles is difficult and
$T_c's$ are correspondingly small.
Comparison of Fig.~\ref{Tc_atomic} for parameters
consistent with atomic polarizations with Fig.~\ref{Tc_cluster}
for  charge transfer excitations makes the distinction of the two
scenarios evident. In the atomic case, $T_c$ is so small that an
observable effect may be found only for the smallest value of
$r/a$. Moreover, the excitation energy is so high that a decrease
of $T_c$ is not seen, even for the largest electric fields. Of
course, when band filling is above half filling, the transition
temperature will decrease with increasing field. However these
field strengths are beyond electrical breakdown.

\section{conclusions}

We have proposed that the field-induced 2DEG in a field effect
device may become superconducting entirely due to pair
interactions with a proximate insulating layer with high
polarizability.  In the general case, properties of a
DS-channel embedded in an oxide field effect transistor will then
be  controlled not only by the electronic and structural
properties of the DS-channel, as is usually assumed, but can
furthermore be strongly influenced by the gate insulator or by
other adjacent dielectric layers.  The choice of the gate
dielectric layer may then influence the device behavior of a field
effect transistor far beyond controlling the maximum gate
polarization and gate current.

For a concrete model of electrons in a 2D layer interacting
with 2-level systems at the interface with the insulator, we
calculated the superconducting critical
temperature, which displays as a function of applied field a steep
initial rise and subsequent decay.
The rise is caused by the increasing density of charge carriers
swept to the interface by the electric field and the decay
is due to the interaction of the field-induced dipoles with the charge
carriers. The optimal values of the field or gate
voltage, as well as sample dimensions and dielectric properties,
were discussed in some detail within the framework of this model.

The goal of these model calculations was to establish the
plausibility of superconductivity in field-effect devices enhanced
or induced by the presence of a polarizable interface, and to
investigate the magnitude of the critical temperature and
likely dependence on external field and materials
properties.  We recognize, however, that several potentially important aspects of
the physics are not present in the model used.  These include
long-range Coulomb interactions,  dynamics of the dielectric
screening, and correlations in the 2DEG.  We anticipate that the
effects of long-range Coulomb interactions are less important in
the present case than in 3D metallic superconductors due to the
larger dielectric constants in the insulating material, but this
must clearly be justified by a legitimate calculation of the true
Coulomb pseudopotential  entering the MacMillan formula.
Screening of the bare electron-dipole interaction $V_{\rm sp}$ must
also be included; we expect, however, that the form of the
dependence of $T_c$ on the electric field will not change significantly
(see Fig.~\ref{Tc_cluster}). The inclusion of a repulsive local
interaction $U$ does not qualitatively change this form either.
However it reduces $T_c$ (by a factor of 2 for $U/4t=0.8$) and it
shifts the optimal value of the field-induced doping to lower levels
(cf.\ Fig.~\ref{finite_U}).

If the effective on-site Coulomb repulsion in the drain-source layer 
is much larger than we have considered, s-wave superconductivity will 
be completely supressed.  It is still tempting, however, to regard the starting
Hamiltonian (\ref{Ht})--(\ref{Hmu}) as the hopping of correlated
electrons (as in the $t-J$ model) where Coulomb interactions have
already been accounted for.  In this case, however, one must
implicitly assume that doubly occupied sites have been projected
out, so that non-retarded $s$-wave superconductivity is
impossible.   Modelling superconductivity within this framework in
higher-angular momentum pair channels will require including
interactions among the two-level systems at the interface, so as to
produce a nonlocal pair potential.
While we have not calculated these effects explicitly, as they are
technically significantly more difficult, our expectation is that
the creation and eventual suppression of the paired state by the
electric field will be qualitatively similar to what we have
calculated. This expectation is supported by the observation that
an increase in $T_c$ with doping has to result trivially from an enhanced
carrier density. Moreover, the buildup of a field-dependent repulsive
interaction $V_z$ has to take place due to the eventual saturation of the dipole
moments with electric field, even when the two-level systems interact.
Investigations along these lines are in progress.

In addition, our considerations can in principle be applied to other,
more weakly correlated systems than the copper oxides.
Our model may be used to treat situations in which the DS-channel
contains  an attractive $s$-wave interaction besides the
interface-mediated pairing. In this case we expect a qualitatively
similar behavior with a nonmonotonic $T_c$.

According to our calculations, devices with thickness of the
insulating layer of order 1000~\AA\ subjected to voltages of order
$\sim$~20V could display superconductivity, if an optimal material
can be found with a sufficiently large low frequency dielectric constant
of order 30-100 and strong quasilocalized electronic modes of
energy $\Delta_{sp}/4t \sim 2.5$~\cite{comment6}. With a 20~V gate
field and a typical 18~nF capacitance
of a 0.03~cm$^2$ gate dielectric (1000~\AA), this would correspond
to a surface charge density of roughly 12~$\mu C/{\rm cm}^2$.
It is now routine to fabricate epitaxial high $\epsilon$ oxide
dielectrics, such as SrTiO$_3$ with polarizations in the range of
10-40~$\mu C/{\rm cm}^2$ \cite{mannhartreview}. Interface-mediated
2D superconductivity therefore seems to us to be plausibly within reach
if good interfaces can be manufactured.

{\it Acknowledgements.} This work is supported by NSF-INT-0340536,
DAAD D/03/36760, NSF DMR-9974396, BMBF 13N6918A,
by a grant from the A.~v.~Humboldt foundation, by
the Deutsche Forschungsgemeinschaft through SFB~484, by the
ESF THIOX programme, and by the Texas Center for Superconductivity
at the University of Houston. The authors are grateful to Y.~Barash,
A.~Hebard, P.~Kumar, G.~Logvenov, D.~Maslov, K.A.~M\"uller, T.S.~Nunner,
N.~Pavlenko, C.W.~Schneider and D.~Tanner for useful conversations.
We thank A.~Herrnberger for his help with the preparation of Fig.~$1$.


\begin{thebibliography}{99}


\bibitem{Little} W.A.~Little, Physical Review A{\bf 134}, 1416 (1964).
\bibitem{Ginzburg} V.L.~Ginzburg, Phys. Lett. {\bf 13}, 101 (1964).
\bibitem{ABB} D.~Allender, J.~Bray, and J.~Bardeen, Phys. Rev. B {\bf 7},
             1020 (1973); {\bf 8}, 4433 (1973).
\bibitem{Ginzburg2} V.L.~Ginzburg and  D.A.~Kirzhnits,
                  {\it High Temperature Superconductivity},
                  Consultants Bureau, New York (1982).
\bibitem{Hebard} A.T.~Fiory, A.F.~Hebard, R.H.~Eick, P.M.~Mankiewich, R.E.~Howard,
                 and M.L.O'Malley, Phys. Rev. Lett. {\bf 65}, 3441 (1990);
                 ibid, {\bf 66}, 845 (1991)
\bibitem{mannhart91} J.~Mannhart, J.G.~Bednorz, K.A.~M\"uller, and
D.~G.~Schlom, Z.\ Phys.\ B {\bf 83}, 307 (1991).
\bibitem{mannhart96} J.~Mannhart, Supercond.\ Sci.\ Technol.\ {\bf 9},
49 (1996).
\bibitem{mannhartreview}  See for example: J.~Mannhart, Mod. Phys.
Lett. B {\bf 6}, 555 (1992);
C.H.~Ahn, J.-M.~Triscone, and J.~Mannhart, Nature {\bf 424}, 1015 (2003).
\bibitem{GennaLog} G.~Yu.~Logvenonv, A.~Sawa, C.~W.~Schneider,
                   and J.~Mannhart, Applied Physics Letter {\bf 83}, 3528 (2003); \\
                   Ann. Phys. {\bf 13}, No. 1-2, 66-67 (2004)
\bibitem{Pallecchi01}  I.~Pallecchi, G.~Grassano, D.~Marre, L.~Pellegrino,
                      M.~Putti, and
                      A.S.~Siri, Appl. Phys. Lett. {\bf 78}, 2244 (2001).
\bibitem{Ueno03} K. Ueno,~I.H.~Inoue, H.~Akoh, M.~Kawasaki, Y.~Tokura,
                and H. Takagi, Appl. Phys. Lett., {\bf 83}, 1755 (2003).
\bibitem{Cassinese04} A.~Cassinese, G.M.~De Luca, A.~Prigiobbo, M.~Salluzzo, and R.~Vaglio,
Appl.\ Phys.\ Lett., {\bf 84}, 3933, (2004).
\bibitem{comment7} Alternatively, charges can be introduced from the drain-source channel
in an enhancement mode FET geometry.
\bibitem{Wehrli} S.~Wehrli, D.~Poilblanc, and T.M.~Rice, Eur.\ Phys.\ J.\ B {\bf 23}, 345-350 (2001).
\bibitem{vandenbrink} J.~van den Brink, M.B.J.~Meinders, J.~Lorenzana,
          R.~Eder, and G.A.~Sawatsky, Phys. Rev. Lett. {\bf 75}, 4658 (1995).
\bibitem{su} W.P.~Su, Phys. Rev. B {\bf 67}, 092502 (2002).
\bibitem{pavlenko} N.~Pavlenko and F.~Schwabl,
                  Phys. Rev. B {\bf 67}, 094516 (2003); Applied Phys. A,
                  in press (cond-mat/0309250).
\bibitem{comment1} For the replacement of the pseudo spin operators
to be valid, the ``low-temperature'' condition $E^\star_{2l}/T\gg 1$
has to be satisfied
($E^\star_{2l}$ is defined in Eq.~\ref{en}).
\bibitem{comment2} We point out that even with the linearization of the
                   'spin excitations' the model is not exactly soluble.
\bibitem{comment3} Electron and hole doping both result in a positive sign of $V_{z}$.
\bibitem{zheng} Zheng~Hang, Phys.\ Rev.\ B {\bf 36}, 8736 (1987).
\bibitem{fehske} H.~Fehske, H.~R\"oder, G.~Wellein, and A.~Mistriotis,
    Phys. Rev. B {\bf 51}, 16582 (1995).
\bibitem{roeder} H.~R\"oder, J~Zang, and A.R.~Bishop,
    Phys. Rev. Lett. {\bf 76}, 1356 (1996).
\bibitem{yuan} Q.~Yuan and P.~Thalmeier, Phys.\ Rev.\ Lett.\ {\bf 83},
3502, (1999).
\bibitem{feynman} R.P.~Feynman, {\it Statistical Mechanics},
Addison-Wesley (1972).
\bibitem{comment4} The nonlinear contributions of the
                   Hamiltonian, Eq. (\ref{H_V_x}) -- (\ref{H_V_x_V_z}),
                   are zero in the thermodynamic average
                   with $\langle \cdots \rangle_{\rm test}$.
\bibitem{comment5} In fact, the virtual exchange of delocalized excitons has
                   been investigated in the previous literature, for example
                   in Refs.~[\onlinecite{Little,ABB}]. The delocalization is not adverse
                   to the formation of the superconducting state.
\bibitem{christen} H.-M.~Christen, J.~Mannhart, E.J.~Williams, and
Ch.~Gerber, Phys.\ Rev.\ B {\bf 49}, 12095 (1994).
\bibitem{comment6} We note that the mechanism we discuss here may also
                   be effective in
                   optimized semiconducting field effect heterostructures,
                   provided the charge carrier density $n$
                   in the interfacial layer can be made of the order of $0.1$.


\end{thebibliography}
\end{document}